\documentclass[12pt]{article}   
\usepackage{epsfig,times}  
\usepackage{color}

\begin{document}
\thispagestyle{empty}

 \renewcommand{\topfraction}{.99}      
 \renewcommand{\bottomfraction}{.99} 
 \renewcommand{\textfraction}{.0}


\newcommand{\nc}{\newcommand}

\nc{\qI}[1]{\section{{#1}}}
\nc{\qA}[1]{\subsection{{#1}}}
\nc{\qun}[1]{\subsubsection{{#1}}}
\nc{\qa}[1]{\paragraph{{#1}}}

\def\qbu{\hfill \par \hskip 6mm $ \bullet $ \hskip 2mm}
\def\qee#1{\hfill \par \hskip 6mm #1 \hskip 2 mm}

\nc{\qfoot}[1]{\footnote{{#1}}}
\def\qL{\hfill \break}
\def\qpar{\vskip 2mm plus 0.2mm minus 0.2mm}
\def\tvi{\vrule height 12pt depth 5pt width 0pt}
\def\qtvi{\vrule height 2pt depth 5pt width 0pt}
\def\qth{\vrule height 12pt depth 0pt width 0pt}
\def\qtb{\vrule height 0pt depth 5pt width 0pt}

\def\qparr{ \vskip 1.0mm plus 0.2mm minus 0.2mm \hangindent=10mm
\hangafter=1}

\def\qdec#1{\par {\leftskip=2cm {#1} \par}}

\def\qdpt{\partial_t}
\def\qdpx{\partial_x}
\def\qddpt{\partial^{2}_{t^2}}
\def\qddpx{\partial^{2}_{x^2}}
\def\qn#1{\eqno \hbox{(#1)}}
\def\qds{\displaystyle}
\def\qw{\widetilde}
\def\qmax{\mathop{\rm Max}}   
\def\qmin{\mathop{\rm Min}}   

\baselineskip=25pt      


\def\qci#1{\parindent=0mm \par \small \parshape=1 1cm 15cm  #1 \par
               \normalsize}

\null
\vskip 2mm

\centerline{\bf \Large A physicist's view of the notion of ``racism''}

\vskip 1cm
\centerline{Charles Jego $ ^{\# 1} $,\quad  Bertrand M. Roehner $ ^{+2} $ }
\vskip 3mm
\centerline{\#: Ecole Polytechnique,\quad +: Institute for 
Theoretical and High Energy Physics}

\vskip 15mm

{\bf Abstract}\quad 
It is not uncommon (e.g. in the media)
that specific groups are categorized as being racist. 
Based on an extensive dataset of intermarriage statistics
our study questions the
legitimacy of such characterizations. It suggests that, far from being 
group-dependent, segregation mechanisms are instead situation-dependent.
More precisely, the degree of integration of a minority in terms of 
the frequency of 
intermarriage is seen to crucially depend upon the 
the proportion $ p $ of the minority. Thus, a
population may have
a segregative behavior with respect to a high-$ p $ ($ p > 20\% $) minority $ A $ 
and 
at the same time a tolerant attitude toward a low-$ p $ ($ p<2\% $) minority $ B $.
This remains true even when $ A $ and $ B $ represent the same minority;
for instance Black-White intermarriage is much more frequent in 
Montana than it is in South Carolina. In short,
the nature of minority groups is largely irrelevant, the key factor being their
proportion in a given area.

\vskip 5mm
\centerline{April 21, 2007}


\vskip 1cm
Key-words: intermarriage, minority, ethnically mixed couples, integration, 
segregation, interaction.
\vskip 1cm 

1: Charles Jego, Centre de Physique Th\'eorique, 
Ecole polytechnique, CNRS, 91128 Palaiseau, France
\qL
\phantom{1: }E-mail: jego@clipper.ens.fr
\vskip 5mm

2: Bertrand Roehner, LPTHE, University of Paris 6, 4 place Jussieu, 
F-75005 Paris, France.
\qL
\phantom{2: }E-mail: roehner@lpthe.jussieu.fr
\qL

\vfill \eject

\qI{Introduction}

In the {\it New York Times} of 24 February 1980 one reads the following
title ``Swedes discover their dark side: racism''. This is by no means
an isolated example; the medias frequently apply
the terms ``racism'' or ``racist'' to populations or peoples. 
Over the period 1971-2005 {\it New York Times} articles featuring these words 
appeared  with a frequency of  57 articles per year. 
For a scientist this raises
the question of how these notions can be defined objectively and whether it is
legitimate to apply them to groups of people or even to whole
nations as in the example above. Naturally, it is
well known that there is no scientific definition whatsoever
of the concept of race, but one can rely on the 
self-identification definition used in U.S. censuses.
Through that procedure one can define (at least for statistical purposes)
populations and groups composed of
``Whites''
\qfoot{Throughout this paper, ``White'' (W) means ``White non Hispanic''.}
, 
``Blacks'' (or ``Afro-Americans''), ``American-Indians'' and so on. 
A commonly held belief is that American states belonging to
the Deep South (Alabama, Louisiana, Mississippi, etc.) are more ``racist''
than northern states. Can such a claim be supported by quantitative evidence
in a way which is consistent with the {\it ceteris paribus} 
(i.e. ``all other things being equal'') requirement?
Econophysics was founded on
the claim that ideas from physics can help us understand social phenomena.
This paper hopes to be an illustration of this claim.
\qpar

In the social science literature
the question of segregation 
is most often considered from an anthropomorphic
perspective, by which we mean that most studies single out
specific populations and rely on factors such as religion, socioeconomic
status, dating circumstances and so on
(Clark-Ibanez et al. 2004, Houston et al. 2004,
Kalminjn 1998, Pagnini et al. 1990. Tatum 1997). 
In contrast, from a physicist's perspective the interaction between ethnic 
groups is naturally seen as 
a case of forming bonds between two types of units, a point of view
which naturally leads to comparative investigations. It is fair to say
that the comparative perspective was also adopted by some sociologists such as
Blau et al. (1984), Duncan et al. (1959), Lieberson et al. (1959), although 
it was not developed in a systematic way. 
\qpar

To summarize the gist of our argument 
by a quick example let us consider the case of Louisiana. 
This state belongs to the
Deep South belt which until the mid-1960s had a well established tradition
of segregation; moreover the Katrina disaster of 2005 revealed 
that inter-ethnic tension between Blacks and Whites is
just beneath the surface.
Apart from its substantial Black minority, Louisiana
also has a small minority of American Indians.
But whereas it
has one of the smallest proportions%
\qfoot{In a normalized sense which will be explained below.}
of Black-White (B-W) 
couples among all US states,
it has one of the
largest proportions of American Indian - White (I-W) couples. 
This example suggests that
speaking of Louisiana Whites as being a group prone to segregationist
attitudes without further qualification
is not consistent with observation. 
The low rate of B-W intermarriage in Louisiana is mainly brought about by
the fact that
Blacks represent a proportion of 32\% in the total population, whereas
the proportion of American Indians is only 0.56\%. 
Naturally, this effect is by no means specific to the United States. 
Back in 1893, people of Italian descent made up 17\% of the
population of Marseilles in the south of France and it can be recalled
that on 16-17 August serious clashes between
French and Italian workers near Aigues-Mortes resulted in the death
of 18 people%
\qfoot{More details can be found in Roehner 2004, p. 197-198.}%
.
\qpar

There are several ways of defining ethnic segregation/integration quantitatively,
namely: (i) Residential integration (ii) School integration (iii) Marriage integration
(iv) Economic integration. 
The first two criteria are closely related for the obvious reason that residential
segregation at block or county level results in {\it de facto} 
school segregation simply because pupils attend school in the area in which 
they live. Residential segregation has been measured by several sociologists%
\qfoot{See for instance: Sorensen et al. (1975), Lieberson (1980),
and Iceland et al. (2002).}
while the second and third criteria have been less studied.
In the present paper, we use the criterion of marriage integration. 
The conclusions drawn from this criterion are to a large extent
consistent with results based on residential segregation  and
school integration (more on this below).
One advantage of the inter-marriage criterion is that one would expect it to be
less dependent on economic
conditions than the residential criterion because it seems possible for
two persons to meet one another (and possibly to get married) even if they
live in segregated areas; workplaces, dance halls, stadiums, holiday resorts
provide contact opportunities which
to some extent are independent of housing location
(see Houston et al. 2005). The fourth criterion would lead us to consider
segregation in the jobmarket and workforce. As census data contain much 
information on occupations they would allow us to carry out such an 
investigation but we will leave it to a subsequent paper.
\qpar

The paper is organized as follows.
First we explain the methodology and test it on what we
call a ``null-experiment''. Then we describe our results for
ethnically-mixed couples. 

\qI{Methodology}

Individual microdata from American censuses are available online on
the website of the Minnesota Population Center%
\qfoot{http://usa.ipums.org/uta/redirect-landing.shtml; IPUMS means
Integrated Public Use Microdata Series.
From a practical perspective, it must be
emphasized that the IPUMS website is very user friendly in the sense that
the steps of selecting, downloading, uncompressing and reading
the data through an appropriate software can be performed within a few minutes.}%
.
Fifteen federal censuses ranging from 1850 to 2000 are accessible through
1\% samples; in addition, 5\% samples are available for some years%
\qfoot{While most of the samples are composed of randomly selected 
households, some samples are not random in the sense that specific 
categories are over-represented. In the present study we used only
random samples.}%
.
Once the data have been selected, 
we count the 
number of ethnically-mixed couples in each state. For instance, using an
unweighted random 1\% sample
of the 2000 census 
we find 3,400 Black-White
couples in Alabama and 400 in New Hampshire.
These counts include married couples 
(identification code 0201) as well
as unmarried male-female partners (identification code  1114).
To be compared in a meaningful way, these numbers must be normalized in
two ways. 
\qbu A first natural normalization is to
compute the number of mixed couples with respect to total number of married
couples.
In 2000 Alabama and New Hampshire had 0.906 million and 
0.262 million married couples respectively.
Thus, one obtains proportions of $ 3,753 $ and $ 1,527 $ B-W couples
per million couples respectively.
However,
this comparison is still meaningless because it fails to take into account
the respective numbers of Black people in each state, namely
259,000 in Alabama versus
7,300 in New Hampshire. To take this difference into account we need a second
normalization. 
\qbu Let us denote by $ p $ the proportion of a minority $ B $ in a population $ A $.
Then, it can be shown by a combinatorial argument 
that if
male-female pairs  are formed randomly in a population of size $ n $ the expected 
proportion of mixed couples (for large $ n $) is:
$$ e_{A-B}= 2p(1-p)  \qn{1} $$ 
It can be noticed that 
if $ p=0.5 $  formula (1) gives $ e_{A-B}=0.5 $ as expected. The case
$ p \ll 1 $ which corresponds to a small population immersed in a much larger
population is of special interest because
it corresponds to most of the minorities to be found
in the United States (American Indians, Chinese, Japanese); in this
case, (1) leads to $ e_{A-B} \simeq 2p $. 
Two crucial assumptions are made in the derivation of (1): 
(i) selection of husband and wife occurs randomly which means 
in particular that it is not subject to any distance limitation; in other words
the probability of a marriage is the same whether both people live
in southern California or in different parts of California. 
(ii) there are no 
institutional or social restrictions in the pairing of $ A $ and $ B $ people.
In real life, these assumptions are usually not fulfilled. Indeed, because of
housing segregation, the vicinity of $ B $ individuals comprises a 
proportion of $ B $ people which may be much larger than the proportion
in the total population. Secondly, even once $ A-B $ contacts have been
established, marriage may not follow due to the ``barrier''
of social conventions. As a result of these restrictions, actual 
rates of mixed couples show a discrepancy with respect
to the rate given by (1) and the magnitude of this discrepancy can serve
to measure the lack of integration. In short, the rationale 
of our normalization procedure is that
equation (1) will be used not as a model but as a yardstick.
\qpar

The normalization procedure can be summarized through the following 
formula giving the normalized frequency $ f_{A-B}(S) $ of $ A-B $ couples 
in state $ S $:
$$ f_{A-B}(S)= { c_{A-B} \over C }{ 1 \over 2p(1-p)}       \qn{2} $$

where
$ c_{A-B}= $ number of mixed couples living in state $ S $, $ C=$ number of
married couples living in state $ S $ and $ p= $ proportion  of the minority
$ B $ in the total population of state $ S $.
$ f_{A-B}(S) $ defines a propensity for integration through marriage.
For the sake of brevity, we 
subsequently refer to it as a marriage integration index
and express it in percent. In a perfectly
integrated society $ f_{A-B} $ would be equal to 100\%, as illustrated by
the case of people born in California considered below.
In a society with a strong propensity for endogamy, $ f_{A-B} $ will be
much smaller; on the contrary, in 
a society with
a strong inclination for exogamy, $ f_{A-B} $ will be larger than 100\% %
\qfoot{The writings of ethologists , ethnologists and anthropologists 
(e.g. Frazer 1910, Karandikar 1929, Makarius 1961, Kortmulder 1968)
suggest the existence of a ``natural'' tendency to exogamy.
If this
were not the case the very notion of species would have little meaning; for
instance elephants of Eastern Africa would in time have become markedly different
from those of Central Africa. In an other context, the tradition for royal heirs to
contract exogamic marriages by taking their wives in foreign courts which probably
goes beyond the desire to establish diplomatic ties. 
On the contrary, diversification within a given
species can be created 
artificially by suppressing interbreeding; in this way 
large numbers of different races can be generated in a relatively short
time span and maintained as long as isolation is enforced, a process
illustrated by the numerous (over one hundred) races of cats and dogs,
all of which are
characterized by markedly different phenotypes. Diversification can also
arise slowly by genetic drift in situations of isolation 
as for instance on islands. }%
. 
\qpar

Returning to our previous example and noting that
in Alabama, $ p=26\%$
whereas in New Hampshire $ p=0.73\% $ one gets
expected proportions $ e_{W-B}=38\% $ and $ e_{W-B}=1.4\% $ respectively. 
Thus,
the B-W marriage integration indexes are
$ f_{W-B}(\hbox{\small Alabama})=0.37/38=0.010 $  
and $ f_{W-B}(\hbox{\small New Hampshire})=0.15/1.4=0.11 $, an integration index
that is about 10 times larger than in Alabama.
\qpar

{\bf Null experiment}\quad 
Before giving complete results for all 50 states we wish to test the 
normalization procedure
through a ``null-experiment'', by which we mean a test-observation
of a situation in which
one does not expect any segregation effect.
To this end,
we consider the minority formed in all states (except California)
by the people who were born in California%
\qfoot{This state was selected because it has the largest population, but
similar tests carried out for New York State and Illinois
led to comparable results.}%
.
In addition, in order to eliminate all effects that may be related to ethnicity
we restrict the sample to White non Hispanic people. In this experiment we count
as mixed couples any couple in which only one of the spouses is born in
California. The results are summarized in Fig. 1. 
  \begin{figure}[tb]
    \centerline{\psfig{width=12cm,figure=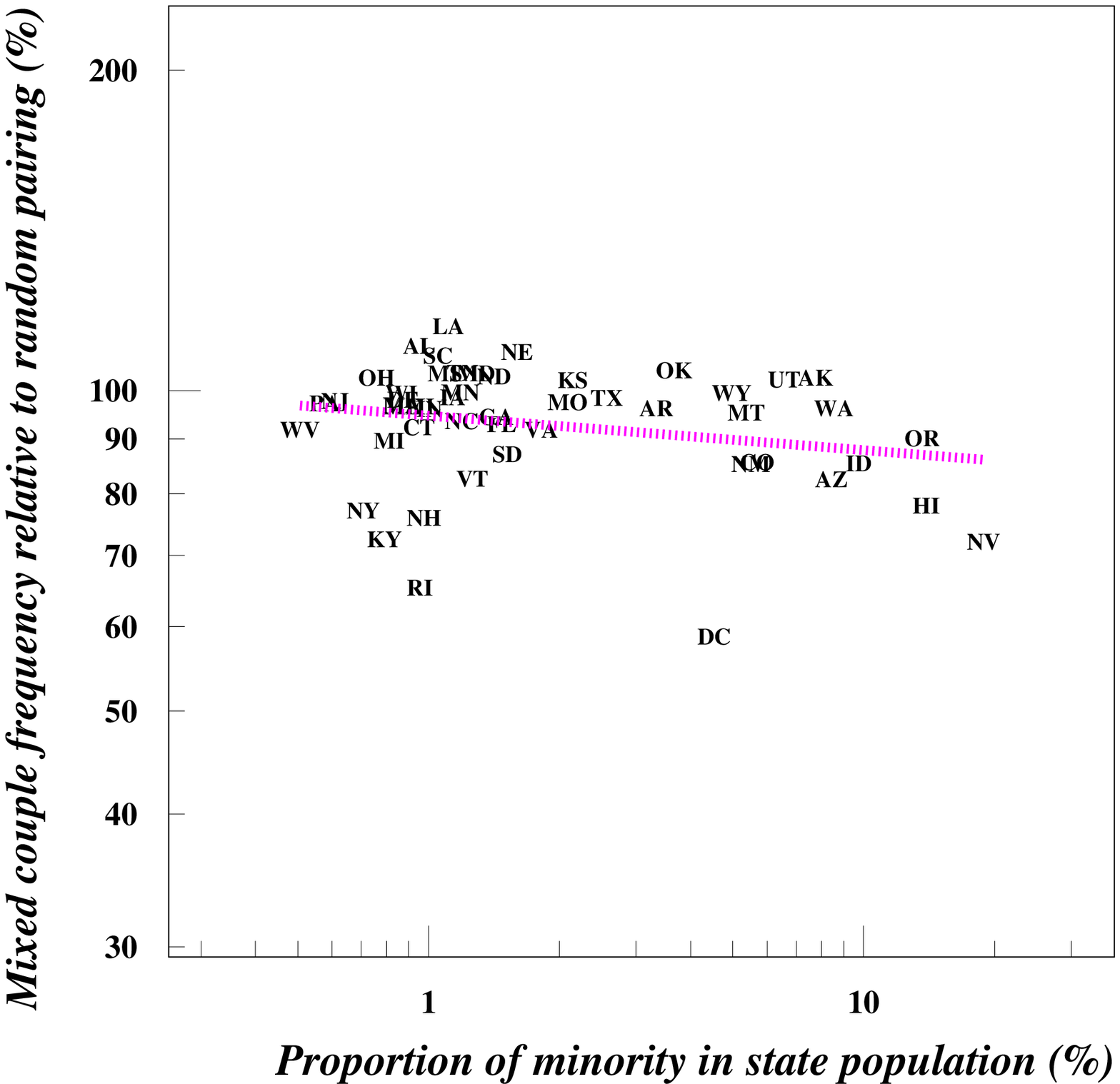}}
\vskip 2mm
    {\bf Fig. 1: Normalized frequency of couples (born in Cal.) -- (not born in Cal.), 
2000 Census.} 
{\small Horizontal scale: percentage of residents of the state who were 
born in California (thereafter called the minority).
Vertical scale: frequency of couples in which one of the partners
(not both) belongs to the minority; this frequency has been normalized
with respect to the number of married couples in
the state and with respect to the importance
of the minority (details can be found in the text in the paragraph about the
definition of the marriage integration index). 
Each label refers to one of the states
plus Washington, DC (California has been excluded). The sample has been
restricted to White non Hispanic people. 
This is a situation in which
one does not expect any ethnic segregation effect in other words one expects
an horizontal scatter plot
at a level close to 100\% , which is indeed what is observed.
The slope of the
linear regression is $ 0.032\pm 0.04 $. Similar results hold for couples in
which one of the partners is born in Illinois or in New York State.}
{\small \it Source: The data are from a $ 1\% $, random
sample of the 2000 Census, available online on the website of the
Minnesota Population Center.}
 \end{figure}
The graph suggest
two comments: (i) For most of the states, the frequency of mixed couples is
close to 100\% which  is in conformity with randomly formed pairs. 
(ii) As expected on account of the lack of ethnic identification, the
slope of the regression line is consistent with a zero value, 
$ a=-0.032 \pm 0.04 $.

\qI{Inter-marriage}
We now repeat the previous procedure for ethnically mixed couples. Fig. 2a
corresponds to the case of Black-White (B-W)
couples; it shows that the frequency
of mixed couples is at least 10 times smaller than in Fig. 1. 
  \begin{figure}[tb]
    \centerline{\psfig{width=12cm,figure=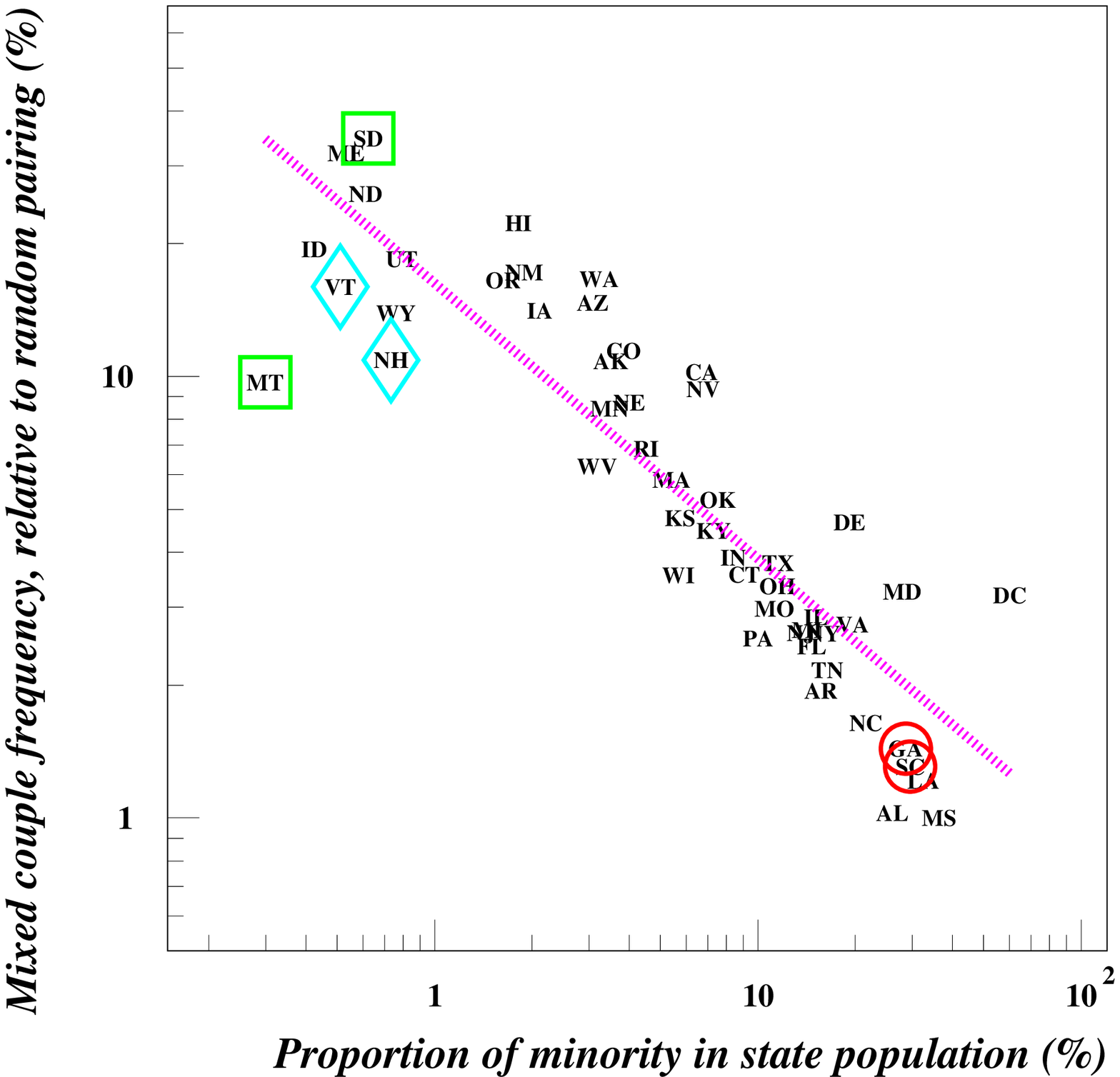}}
\vskip 2mm
    {\bf Fig. 2a: Normalized frequency of Black -- White couples, 2000 Census.} 
{\small Horizontal scale: percentage of the Black population.
Each label refers to one of the states plus Washington, DC. Circles are
drawn around Georgia and South Carolina (states with high proportions of Blacks),
squares are drawn around Montana and South Dakota (states with low 
proportions of Blacks), diamonds are drawn around New Hampshire and
Vermont (states with low proportions of Blacks).
The correlation is $ -0.89 $ and the slope of the
linear regression is $ -0.62\pm 0.09 $}
{\small \it Source: The data are taken from an unweighted $ 1\% $
sample of the 2000 Census.}
 \end{figure}
The
frequency of B-W couples has tripled in the period 1970-2000 but it still
remains at a low level. In addition, there is a marked negative slope
$ a=-0.62\pm 0.09 $. The pattern for American Indian - White (I-W) couples is
similar but the frequency is about 4 times higher and the slope about one half
of the previous one: $ a=-0.36\pm 0.12 $, see Fig. 2b. 
  \begin{figure}[tb]
    \centerline{\psfig{width=12cm,figure=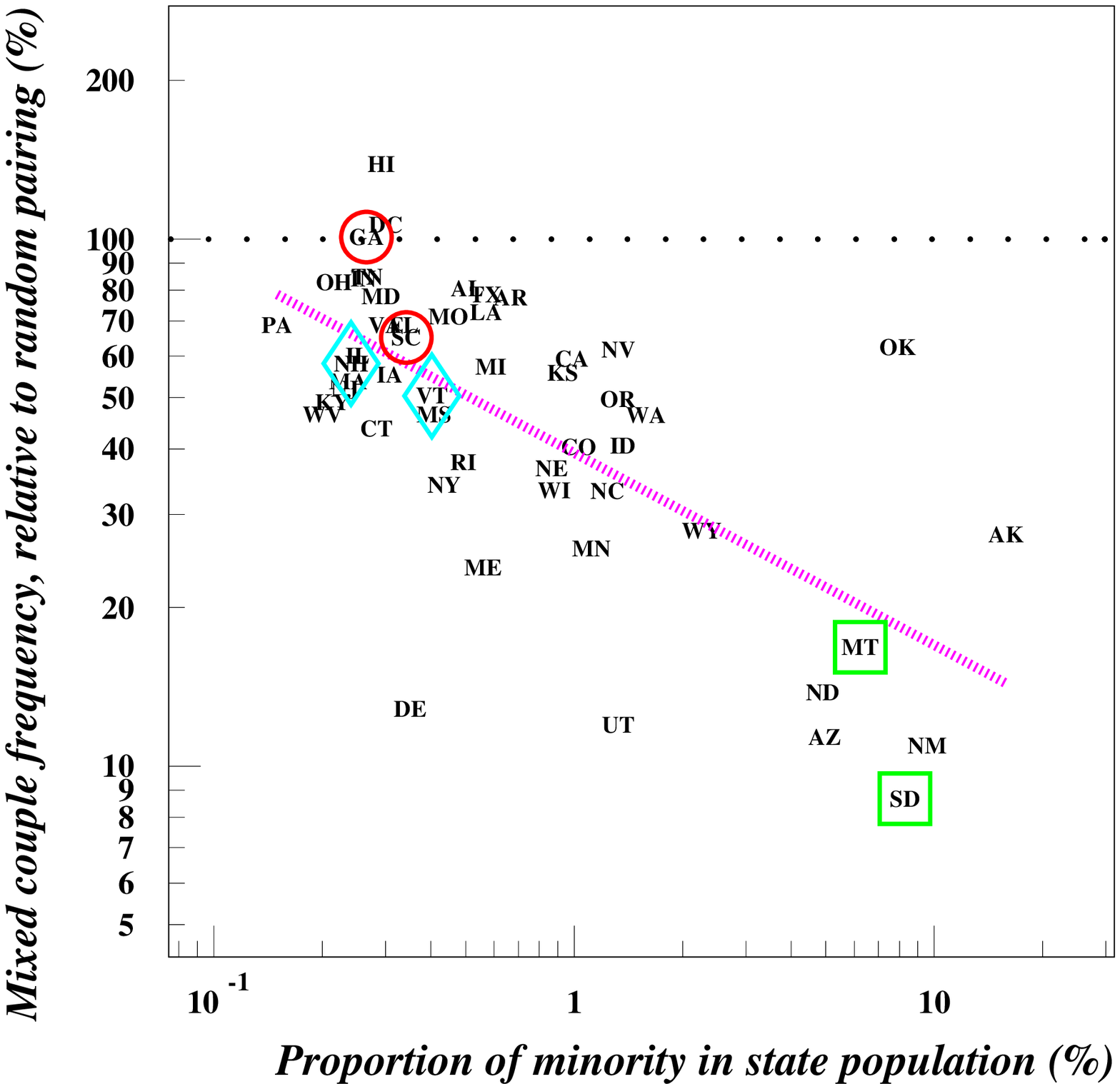}}
\vskip 2mm
    {\bf Fig. 2b: Normalized frequency
 of American-Indian -- White couples, 2000 Census.} 
{\small Horizontal scale: percentage of the American Indian population 
with respect to total state population.
It can be seen that Georgia and South Carolina (circled) which had a small
mixed couple frequency in Fig. 2a have a high frequency here, 
whereas Montana and
South Dakota (squares) 
which had a high frequency in Fig. 2a have a low one here due to their
substantial proportion of American Indians.
Vermont and New Hampshire (diamonds) which
have only few minority residents of both kinds have a high frequency in 
both graphs. Similar conclusions can be drawn as well from the data
for other states.
The correlation is $ -0.66 $
(confidence interval for probability $ 0.95 $ is $ -0.79 $ to $ -0.47 $);
the slope of the regression line is $ -0.36\pm 0.12 $.}
{\small \it Source: The data are taken from an unweighted $ 1\% $, 
sample of the 2000 Census.}
 \end{figure}
\qpar

The most interesting observation
is the fact that the states of the Deep South (e.g. Alabama, Arkansas, Louisiana,
South Carolina) which have low B-W frequencies are at top frequency levels
in Fig. 2b. Similarly, states such as Arizona, Montana or South Dakota
have low I-W frequencies but high B-W frequencies. In addition
states such as New Hampshire or Vermont which do not have any substantial
ethnic minority whatsoever are at the top of the scatter plots in both graphs. 
\qpar

The observation that low 
minority percentages are associated with higher integration levels
can be made
for other minorities as well. Let us give some examples.
\qbu Alabama, Arkansas and Florida have small percentages 
($ p_{\hbox{J}}< 0.1 \% $) 
of Japanese Americans and some of the highest frequencies of
Japanese Americans - White (J-W) couples. In contrast, Hawaii has both
the highest percentage of ethnic Japanese 
($ p{\hbox{J}} \simeq 10\% $) and the lowest
frequency of J-W couples; there is a  ratio of about 10 between the frequencies
in Florida and Hawaii. 
\qbu The frequency of mixed Hispanic - non Hispanic couples is about 
three times higher in Louisiana 
($ p_{\hbox{Hisp}}\simeq 2\% $) than in Texas or
California ($ p_{\hbox{Hisp}}\simeq 20\% $). 
\qpar

\qI{Other integration characterizations}

\qA{Alternative criterion}
The previous observations can be confirmed by 
using an alternative criterion which 
does not require a renormalization procedure (at least for small $ p $). We
gauge the exogamous versus endogamous character of a minority by the 
ratio:
$$ \Gamma _A=\hbox{Number of exogamous couples A-X}/\hbox{Number of
endogamous couples A-A} $$

The notation $ X $ instead of $ B $ (as above) refers to the fact
that in this definition {\it all} exogamous couples of $ A $ with any other group
are summed up in the numerator. 
Typical orders of magnitude of $ \Gamma $ are given in Table 1.

\begin{table}[tb]

 \small 

\centerline{\bf Table 1\quad Variation of the exogenous/endogenous ratio
with respect to minority percentage}

\vskip 3mm
\hrule
\vskip 0.5mm
\hrule
\vskip 2mm

$$ \matrix{
 \hbox{}  &  \hbox{Black} &  \Gamma_{\hbox{\small Black}} & \qquad &
 \hbox{Am. Ind.} &  \Gamma_{\hbox{\small Am. Ind.}} \cr
\qtb
 \hbox{}  & \hbox{\%} &   &  &\hbox{\%} &  \cr
\noalign{\hrule}
\qth 
\hbox{\color{green} Montana+South Dakota} \hfill & {\color{green}
0.41} & {\color{green} 1.2\phantom{00} } & &{\color{green} 7.2\phantom{0} } & 
{\color{green}0.79} \cr
\hbox{\color{red} Georgia+South Carolina}  \hfill& {\color{red} 29\phantom{.00} }& 
{\color{red} 0.010 } & &{\color{red} 0.28 }& 
{\color{red} 5.1\phantom{0} } \cr
\hbox{}  \hfill &  &  &  &  \cr
\qtb
\hbox{\color{blue} Vermont}  \hfill & {\color{blue} 0.49} & 
{\color{blue} 1.0\phantom{00} }& & {\color{blue} 0.34 }& 
{\color{blue}7.0\phantom{0} }\cr
\noalign{\hrule}
} $$

\vskip 1.5mm
Notes: A large $ \Gamma $ indicates a high degree of integration whereas
a $ \Gamma $ close to zero suggests a high level of segregation. 
For each minority the first column gives its population percentage.
The first two lines correspond to two contrasting situations in term of
minority proportion; Montana+South Dakota (these states have been
lumped together to increase the number of marriages)
has a sizable proportion of American
Indians whereas Georgia+South Carolina has a substantial Black population. 
In the case of Vermont both minorities have a small
percentage. The table suggests that integration decreases when
the population percentage of the minority increases. It is particularly 
striking that the integration of the Black populations in Montana+South Dakota
and in Vermont is higher than the integration of American Indians in
Montana+South Dakota because usually the integration of  the Afro-American
population 
is fairly low due to a long historical legacy of segregation. 
\qL
Source: The data are from a $ 5\% $ random sample of the 1980 Census.
\vskip 2mm

\hrule
\vskip 0.5mm
\hrule

\normalsize

\end{table}

It shows that:
 $$ \Gamma _{\hbox{\small Am.Ind.}}(\hbox{\small Low proportion of minority}) =6.4\
\Gamma _{\hbox{\small Am.Ind.}}(\hbox{\small High proportion of minority}) $$
 $$ \Gamma _{\hbox{Black}}(\hbox{Low proportion of minority})=118\
\Gamma _{\hbox{Black}}(\hbox{High proportion of minority}) $$

Note that the factors $ 6.4 $ and $ 118 $ cannot be really compared because
what we
call a ``high proportion'' is not the same in the two cases: for
American Indians ``high'' means $ 7.2 \% $, whereas for Blacks it means
$ 29 \% $. In addition there may be reinforcing and cumulating effects
due to high proportions persisting over long periods of time;
this historical aspect we leave for a subsequent study.

\qA{Residential segregation}
At the beginning of the paper we said that our findings are consistent
with observations based on residential segregation. Let us shortly illustrate
this statement by a few examples based on a study published
by the Bureau of the Census (Iceland et al. 2002):
\qbu The most segregated
Metropolitan Area for Blacks in 2000 was Milwaukee-Waukesha in Wisconsin
(segregation index $ \delta=0.89 $
\qfoot{More precisely, this segregation index is related to the fraction
of the Black population that would have to move across blocks
to achieve an uniform
minority density.}
) 
and it had a Black population percentage of $ p=25\% $;
the least segregated Metropolitan Area for Blacks was Orange county
in California ($ \delta =0.52 $) with a Black population representing
$ p=2.0\% $.
\qbu For Asians and Pacific Islanders the most segregated Metropolitan
Area was San Francisco (California): 
$ \delta =0.83,\ p=33\% $ whereas the least segregated
was the Nassau-Suffolk area (New York): $ \delta=0.55,\ p=4.0\% $.
\qpar

\qA{School integration}
The third characterization of ethnic integration that we mentioned is 
school integration. In the late 1950s and early 1960s the {\it New York Times}
published annual maps showing the (fairly slow) progress of school integration.
For instance the map published on 12 May 1963 shows that
the percentage of Black pupils who were in class with Whites was 
close to zero ($ <0.6\% $) in 7 states of the Deep South (Alabama, Arkansas,
Florida, Georgia, Louisiana, Mississippi, South Carolina); moreover, for the
16 Southern states for which the {\it New York Times} gives data
there is a significant correlation ($ r=0.82 $)
between the lack of school integration and the 
proportion of the Black population. 
\qpar

\qA{Hate crimes}
Is racial violence in the form of what the Federal Bureau of Investigation
calls {\it hate crimes} directed against minority members
also increasing with the minority's proportion? As hate crimes are a form
of rejection one would expect that their frequency decreases for any given
minority as this minority becomes better integrated. Such a
relaxation process suggests
that the historical background is of importance. That is why we restrict
our comparison to two communities which have been present in the United
States at least since the end of the War of Independence, namely Blacks
and American Indians. 
In 2000 there were 104 hate crimes against Blacks
per million of their population as compared to a rate of 27  against American
Indians%
\qfoot{Statistical Abstract of the United States 2000, p. 188.}%
. 
These figures are consistent with our previous finding that marriage integration
is substantially higher for American Indians.

\qI{Conclusion}
Using an analysis based on the number of intermarriages in the United States
we have seen that the proportion of minorities in the total population is
a key parameter in order to understand segregation patterns. In the light
of this finding
the title of the {\it New York Times} article
mentioned in the introduction can now be reinterpreted.
Did Swedes really
reveal a facet of their nature 
which had not been apparent so far? One should recall that prior to 1980
there were almost no sizable
ethnic minorities in Sweden; even in 2006 they represented
less than  $ 5\% $ of the population.
Thus, Sweden was in a situation similar to New Hampshire
or Vermont where tolerance is a natural consequence of small values of $ p $.
As $ p $ increased, Sweden faced the kind of situations 
experienced by 
U.S. states with comparable $ p $ values in Fig. 2a,b.
Thus, it is not surprising to see that Swedes reacted more
or less in the same way as residents of those states.
\qpar

{\bf Acknowledgments} Many thanks to Peter Richmond who
offered useful comments and attracted our attention to two attempts in
the same direction by J\"urgen Mimkes (2006, no date).

\vfill\eject


\vfill\eject

{\bf \large References}

\qparr
Blau (P.), Becker (C.), Fitzpatrick (K.) 1984: Intersecting social affiliations and
intermarriage.
Social Forces 62, 585-606.

\qparr
Clark-Ibanez (M.), Felmlee (D.) 2004: Interethnic relationships: the role of
social network diversity.
Journal of Marriage and Family 66, 293-305.

\qparr
Ducan (O.D.), Lieberson (S.) 1959: Ethnic segregation and assimilation.
American Journal of Sociology 64, 4, 364-374.

\qparr
Frazer (J.G.) 1910: Totemism and exogamy: a treatise on certain early
forms of superstition and society. Macmillan, London.

\qparr
Houston (S.), Wright (R.), Ellis (M.), Holloway (S.), Hudson (M.) 2005: Places
of possibility: where mixed-race partners meet.
Progress in Human Geography 29, 6, 700-717.

\qparr
Iceland (J.), Weinberg (D.H.), Steinmetz (E.) 2002: Racial and ethnic residential
segregation in the United States: 1980-2000. 
U.S. Census Bureau. Census 2000 Special Report.

\qparr
Jacobs (J.), Labov (T.) 2002: Gender differentials in intermarriage among
sixteen race and ethnic groups.
Sociological Forum 17, 621-646. 

\qparr
Kalmijn (M.) 1998: Intermarriage and homogamy: causes, patterns, trends.
Annual Review of Sociology 24, 395-421.

\qparr
Karandikar (S.V.) 1929: Hindu exogamy. D.B. Taraporevala, Bombay. 

\qparr
Kortmulder (K.) 1968: An ethological theory of incest taboo and exogamy:
with special reference to the views of Claude Levi-Strauss. 
Current Anthropology 9, 5, 437-449.

\qparr
Lieberson (S.) 1980: A piece of the pie: Blacks and White immigrants since 
1880. University of California Press. Berkeley.

\qparr
Lieberson (S.), Waters (M.C.) 1988: From many strands: ethnic and
racial groups in contemporary America.
Russel Sage Foundation, New York.

\qparr
Makarius (R.) 1961: L'origine de l'exogamie et du tot\'emisme. Gallimard, Paris.

\qparr
Mimkes (J.) 2006: A thermodynamic formulation of social science. In
``Econophysics and sociophysics: trends and perspectives''
edited by B.K. Chakrabarti, A. Chakraborti, and A. Chaterjee, p. 279-310.
Wiley-VCH, Weinheim.

\qparr
Mimkes (J.) no date: Intermarriage as ``thermometer'' of social systems.
Unpublished, personal communication. 

\qparr
Pagnini (D.L.), Morgan (S.P.) 1990: Intermarriage and social distance
among U.S. immigrants at the turn of the century.
American Journal of Sociology 96, 2, 405-432.

\qparr
Peach (C.) 1980: Ethnic segregation and intermarriage.
Annals of the Association of American Geographers 70, 371-381.

\qparr
Roehner (B.) 2004: Coh\'esion sociale. 
Odile Jacob. Paris.

\qparr
Roehner (B.M.) 2007: Driving forces of physical, biological and social
systems. A network theory investigation of social bonds and interactions.
To appear in July 2007. Cambridge University Press. Cambridge.

\qparr
Sorensen (A.), Taeuber (K.E.), Hollingworth (L.J. Jr.) 1975: Indexes of 
racial residential segregation for 109 cities in the United States 1940 to 1970.
Sociological Focus 8, 128-130. 

\qparr
Tatum (B.) 1997: ``Why are all the Black kids sitting together in the cafeteria?''
and other conversations about race.
Basic Books. New York.

\end{document}